# Isolation and Characterization of Few-layer Manganese Thiophosphite

*Gen Long[1], Ting Zhang[1,2], Xiangbin Cai[1], Jin Hu[3], Chang-woo Cho[1], Shuigang Xu[1,4], Junying Shen[1], Zefei Wu[1], Tianyi Han[1], Jiangxiazi Lin[1], Jingwei Wang[1], Yuan Cai[1], Rolf Lortz[1], Zhiqiang Mao[3], and Ning Wang[1,\*]*

**Affiliations:**

*1.    Department of Physics and Center for Quantum Materials, The Hong Kong University of Science and Technology,  Hong Kong, China*

*2.    Institute for Advanced Study, The Hong Kong University of Science and Technology, Hong Kong, China*

*3.    Department of Physics and Engineering Physics, Tulane University, New Orleans, LA-70118, USA*

*4.    National Graphene Institute, University of Manchester, Manchester M139PL, UK*

*\*Correspondence to: phwang@ust.hk*

**Abstract:**

   This work reports an experimental study on an antiferromagnetic honeycomb lattice of $MnPS_3$ that couples the valley degree of freedom to a macroscopic antiferromagnetic order. The crystal structure of $MnPS_3$ is identified by high-resolution scanning transmission electron microscopy. Layer-dependent angle-resolved polarized Raman fingerprints of the $MnPS_3$ crystal are obtained and the Raman peak at 383 cm$^{-1}$ exhibits 100% polarity. Temperature dependences of anisotropic magnetic susceptibility of $MnPS_3$ crystal are measured in superconducting quantum interference device. Magnetic parameters like effective magnetic moment, and exchange interaction are extracted from the mean field approximation model. Ambipolar electronic transport channels in $MnPS_3$ are realized by the liquid gating technique. The conducting channel of $MnPS_3$ offers a unique platform for exploring the spin/valleytronics and magnetic orders in 2D limitation.

The magnetic order in 2D limitation is of boundless interest not only for fundamental condensed matter studies but also as a potential candidate in numerous technological applications[1-7]. Despite the extensive family of 2D crystals, only a few exhibit intrinsic magnetic orders[8-10]. Therefore, research has been mostly limited to the magnetic orders arising from extrinsic effects, such as vacancies, defects, edges or chemical dopants[11-16]. An emerging 2D crystal group, namely, transition-metal thiophosphite (TMT) ($MPX_3$; M=Fe, Ni, Mn, Cd, Zn, etc.; P=P; X=S, Se, etc.), offers new possibilities because of the suitability of TMT as a platform for exploring novel intrinsic magnetic orders[17-19]. Different transition-metal ions in TMT accumulate different antiferromagnetic orders. $FePS_3$ is best described by the Ising model, $NiPS_3$ by the anisotropic Heisenberg model, and $MnPS_3$ by the isotropic Heisenberg model[5, 17]. The intrinsic degrees of electronic freedom, such as charge and spin, have been broadly explored in the last few decades in electronics and spintronics[20-23]. In recent years, a new electron-valley freedom has drawn much attention because of its immense potential for fundamental studies on new quantum concepts and next-generation electronics[24-26]. This valley freedom is predicted to couple with antiferromagnetic order in $MnPS_3$ because of the latter's antiferromagnetic honeycomb lattice[27]. Coupling the micro-nature to the macro-phenomena renders $MnPS_3$ an ideal playground for exploring novel electronic degrees of freedom. In this work, we perform a systematic optical and electronic transport study of $MnPS_3$ in an atomically thin level. In addition to identifying the crystal structure with high-resolution scanning transmission electron microscopy (HRSTEM), Raman fingerprints of $MnPS_3$ with different thickness were also determined through angle-resolved polarized Raman (ARPR) spectroscopy. Temperature dependence of magnetic susceptibility of $MnPS_3$ is measured by a superconducting quantum interference device (SQUID). Critical magnetic parameters like effect magnetic momentum and exchange interactions are extracted from the mean field approximation (MFA) model. Last, we develop a liquid gating (LG) technique and fabricate $MnPS_3$-based electrical double-layer transistors (EDLTs) to determine its electronic transport properties. Benefiting from the high efficiency of LG[28], an ambipolar conducting channel is observed with carrier mobility ranging from 1 $cm^2 V^{-1} s^{-1}$ to 3 $cm^2 V^{-1} s^{-1}$ in $MnPS_3$.

Top view of monolayer $MnPS_3$ is illustrated in Fig. 1a (Upper left panel). Each $[P_2S_6]^{4-}$ unit is located at the center of a regular hexagon with six corners occupied by $[Mn]^{2+}$. The dumbbell-shaped structure of $[P_2S_6]^{4-}$ is shown in Fig. 1b. All the $[Mn]^{2+}$ are arranged in a honeycomb structure, and each $[Mn]^{2+}$ is surrounded by six sulfur atoms. The zigzag direction of $[Mn]^{2+}$ is defined as the $a$ direction, whereas the armchair direction (perpendicular to $a$) is defined as *b*. The point group of monolayer MnPS$_3$ is assigned to be $\bar{3}\,2/m$. The threefold inverse rotation symmetry results a valley degeneracy at the corner (K point) of the hexagonal first Brillouin zone (BZ). When stacked together along the *c* direction ($\beta = 107.5°$), the atomic layers break the threefold inverse rotation symmetry and render the bulk MnPS$_3$ a monoclinic structure with a point group of 2/m (Fig. 1a, right panel). Given the weak van der Waals interaction between atomic layers, the mono- and few-layered MnPS$_3$ flakes can be mechanically exfoliated from the bulk crystal by scotch tape method[29-30]. Fig. 1c shows a representative micro-optical image of MnPS$_3$, with different layers showing distinct optical contrasts due to light interference. To confirm the thickness of these flakes, we perform atomic force microscopy (AFM) measurement and a thickness *h* of 0.8 nm is obtained for one atomic layer (Fig. 1d), which agrees well with the layer space of 0.68 nm[31].

While exfoliating the bulk crystal, we find that most of (>90%) the flakes exhibit quadrilateral shapes having inner angles of ~60° or 120° (Fig. 1c). To elucidate the mechanism behind these two special angles, we compare between the selected-area electron diffraction (SAED) pattern and its defocused transmitted spot to index the edges of the exfoliated MnPS$_3$ flakes. The MnPS$_3$ flake is kept fixed along the *c* direction (Fig. 2a) and the focus is adjusted until the real space features are visible inside the transmitted spot (Fig. 2a, inset). The two edges forming 60° angle are indexed to be (110) and (020) according to the perpendicular relation between the reciprocal space and real space. The (110) and (020) directions correspond to the zigzag directions of [Mn]$^{2+}$ in the MnPS$_3$ structure. This result can be readily explained by the weakest breaking strength along the zigzag direction of [Mn]$^{2+}$ (Ref: 32). To verify this conclusion, we obtain a high-angle annular dark field (HAADF) HRSTEM image of the same sample along the *c* direction (Fig. 2b). Each bright spot corresponding to a single atom, the three elements are indicated by different colored balls, and primary unit cell is marked by a green-dashed parallelogram. The six-membered ring composed of manganese atoms is clearly visible. The variation in intensity along the dashed lines are displayed in the right bottom inset. The lattice parameters extracted from the image (a=6.08±0.05 nm; b=10.52±0.05 nm) agree with the values measured from neutron scattering experiment[31].

Besides structural identification, we also examine the Raman fingerprints of MnPS$_3$ with different layer numbers. Fig. 3a shows the Raman spectra of MnPS$_3$ with different layers collected at room temperature. Two Raman peaks are apparent (P$_{273}$: 273 cm$^{-1}$; P$_{383}$: 383.6 cm$^{-1}$; the Raman peak at 302 cm$^{-1}$ comes from silicon substrate) in the bulk samples. The intensities of $P_{273}$ and $P_{383}$ decrease dramatically with decreasing layer number and disappear in the monolayer flake (Fig. 3b). The main Raman spectral features for the monolayer, bilayer, and trilayer flakes are summarized below. Monolayer flakes exhibit no observable Raman peaks. When layer number increases to two, both $P_{273}$ and $P_{383}$ emerge, and $P_{383}$ is considerably weaker than $P_{273}$. For three-layer flakes, the two peaks at 273 cm$^{-1}$ and 383 cm$^{-1}$ exhibit similar intensities. The clear dependence of Raman peak intensities on layer number renders the Raman spectrum a reliable criterion for determining the thickness of a few-layer MnPS$_3$ sample.

To probe further into symmetry properties, we perform ARPR spectral measurements on a 10 nm-thick MnPS$_3$ flake at room temperature. The ARPR measurement configuration is shown in the top right inset of Fig. 4a. The *a* direction (green arrow) runs along one of the edges of the exfoliated MnPS$_3$ flakes (zigzag direction of MnPS$_3$), and the *b* direction (blue arrow) is perpendicular to *a* direction. The linear polarization direction of the incident laser (purple arrow) and scattered light (light blue arrow) was illustrated as well. The definitions of angles $\alpha$ and $\theta$ are shown in the inset. Fig. 4a presents the ARPR spectra varying with $\alpha$ when $\theta$ is fixed at 0°. $P_{273}$ remains unchanged with increasing $\alpha$, whereas the peak at 383 cm$^{-1}$ vary periodically with $\alpha$. Moreover, $P_{383}$ reaches its maximum at 0° and 180°, and disappears at around 90°. This pattern demonstrates a 100% polarity of Raman peak at 383 cm$^{-1}$ (Ref: 33-34). We perform a fine $\alpha$ (every 10°) dependent ARPR measurement and the extracted intensities of the two peaks are shown in Fig. 4b. The black solid line implies $I(\alpha) = I(\alpha = 0) * \cos^2(\alpha)$, as expected and measured in varying systems. By contrast, the red solid line confirms the depolarized feature of $P_{273}$[34]. The polarized and depolarized behaviors of Raman peaks demonstrate different symmetry properties of the corresponding phonon modes. Specifically, the polarized behavior of the Raman peak signifies that the peaks arise from totally symmetric variations. Fig. 4c presents the observed ARPR spectra at a few different $\theta$ with $\alpha$ fixed at 0°. The independence of the ARPR spectrum on $\theta$ demonstrates that the crystalline orientation hardly affects the Raman spectrum of MnPS$_3$. The phonon spectrum and corresponding Raman susceptibilities at the center of the first BZ are calculated through the density function perturbation theory (DFPT)[35-38] to assign the Raman modes and phonon frequencies (supporting information). The bulk crystal of MnPS$_3$ belongs to the C2/m symmetry group. The 30 irreducible phonon modes at the first BZ center are expressed as $\Gamma = 8A_g + 6A_u + 7B_g + 9B_u$ and confirmed by the calculation results. Polarized peak at 383 cm$^-$

$^1$ is assigned to the B$_g$ mode, whereas the depolarized peak at 273 cm$^{-1}$ is assigned to the A$_g$ mode according to the symmetry elements of the A$_g$ and B$_g$ modes$^{39}$.

Fig. 5a reveals the temperature dependences of mass magnetic susceptibilities $\chi_m$ of MnPS$_3$ bulk crystal for two directions, in-plane $\chi_{m//}$ and out-plane $\chi_{m\perp}$ magnetic fields, measured in SQUID. When temperature is above 200K, the isotropic $\chi_{m//}$ and $\chi_{m\perp}$ demonstrate a Heisenberg-type magnetic order in MnPS$_3$$^{40}$. In the high temperature paramagnetic phase, $\chi_m$ follows the Curie-Weiss law $\chi_m = C/(T-T_C)$, where $C=2.65\pm0.05\times10^{-4}$ $m^3Kkg^{-1}$ is the Curie constant and $T_C$=-390K is the Curie temperature. In the MFA model, the effective magnetic moment $\mu_{eff}$ of [Mn]$^{2+}$ could be extracted from the Curie constant according to $C = \dfrac{2\mu_0 N \mu_{eff}^2}{3k_B}$, where $\mu_0$ is the vacuum permeability, $N$ is the primitive unit cell number per unit mass (also the number of [Mn]$^{2+}$ pairs per unit mass), k$_B$ denotes the Boltzmann constant$^{41-42}$. The derived $\mu_{eff} = 5.6\,\mu_B$ agrees well with the total magnetic moment of 3d$^5$ electron system in high spin state (S=5/2) $\mu_{3d5} = 5.9\,\mu_B$ $^{17}$. To confirm the obtained electronic state of Mn element in MnPS$_3$, we performed the electron energy-loss spectroscopy (EELS) of few-layer flakes (Fig. 5c). The Mn-L$_{2,3}$ edge energy-loss near-edge structures (ELNES) and the chemical shift agree well with the reported [Mn]$^{2+}$ spectrum$^{43}$. The value of $\mu_{eff}$ demonstrates that manganese is in the form of ions rather than atomic form in MnPS$_3$. As temperature cools down, the Curie-Weiss law fails to describe the behavior of $\chi_m$. An isotropic broad peak of $\chi_m$ at 120K suggests a short range order of [Mn]$^{2+}$ spins in the ab plane. Further cooling down the sample, $\chi_{m\perp}$ sharply decreases to 0 while $\chi_{m//}$ remains essentially constant singling the antiferromagnetic order in MnPS$_3$ with T$_N$=77K. T$_N$ is in consistence with the report value of 78K$^{40}$.

The contrasting behaviors between $\chi_{m\perp}$ and $\chi_{m//}$ agree well with the MFA model. The left (right) panel of Fig. 5d present the magnetic susceptibility measurement configuration with an applied magnetic field $H$ perpendicular (parallel) to the spin orientations. When $H$ is perpendicular to the spin orientations, the system energy density can be expressed as: $U = -\mu_0 \lambda M^2(1-0.5(2\alpha^2)) - 2\mu_0^2 HM\alpha$, where $\lambda$ is the Weiss constant, M=|M$_A$|=|M$_B$| is the magnetization strength for the [Mn]$^{2+}$ with unit orientation, $\alpha$ is the orientation of $M_A$ and $M_B$ caused by $H$. $U$ reaches its minima at $\alpha = \mu_0 H/2\lambda M$, hence the magnetic susceptibility with *H* perpendicular to spin orientations (in-plane magnetic field) $\chi_{//} = \frac{2M\alpha}{H} = \frac{\mu_0}{\lambda}$ is a constant. The observed slight increase of $\chi_{m//}$ when cooling down (Fig. 5a) arises from the formation of spin density wave when temperature is lower than $T_N$[17, 44]. When *H* is parallel to the spin orientations (right panel of Fig. 5d), if $M_A$ and $M_B$ make equal angles with *H*, the magnetic field is not changed and the $\chi_{//}(T=0) = 0$ which agrees well with the measured results. The isotropic characteristics of $\chi_m$ at high temperature as well as the anisotropic characteristics of at low temperature confirms the 2D Heisenberg-type magnetic order with spins easy axis perpendicular to the ab plane in MnPS$_3$ as shown in the inset of Fig. 5a.

Another critical parameter of Heisenberg model, the main exchange interaction *J* is estimated from the high field $\chi_{m\perp}$. Fig. 5b presents the mass magnetization accompanied with the derivative mass magnetic susceptibility defined as $\chi_{md} = \frac{1}{m}\frac{dM}{dH}$ with magnetic field perpendicular to the ab plane at T=5K. The mass magnetization exhibits a spin-flop transition feature at $H_C=3.8\times10^6$ A/m corresponding to the peak of $\chi_{md}$. In a high field region above $5\times10^6$ A/m, the mass magnetization increase with increasing magnetic field linearly and $\chi_{md}^h$ =1.08×10$^{-6}$ m$^3$kg$^{-1}$ is obtained. Using the MFA expression $\chi_{md}^h = -\frac{\mu_0 N(g\mu_B)^2}{2zJ}$, where *g=2.0* is the g-factor, *z* coordination number of [Mn]$^{2+}$ and z=3 for honeycomb lattice, the main exchange interaction is extracted to be *J/k$_B$=-20.4 K*[40]. Alternatively, *J* can also be obtained from Curie temperature T$_C$ through $J = \frac{3k_B T_C}{2zS(S+1)}$, where S=5/2 for a high spin state of 3d5 system. When *T$_C$=-390K*, the value of *J* is extracted to be *J/k$_B$=-22.2K*[17]. The negative exchange interaction signals the antiferromagnetic coupling between two adjacent [Mn]$^{2+}$ ions. The universal main exchange interaction achieved from different methods demonstrates a reliable measurement of the magnetic properties of MnPS$_3$. In other words, the universal exchange interaction demonstrates the validity of bare electron g-factor in MnPS$_3$ system.

From the practical viewpoint, the electronic transport properties of MnPS$_3$ flakes are probed. A large MnPS$_3$ band gap of more than 3 eV disables chemical potential tuning between the conduction and valence bands by using a field effect with SiO$_2$ or boron nitride as a dielectric layer[4]. Hence, an LG technique is applied to fabricate the EDLTs, which exhibit two orders of increased efficiency in tuning the carrier density with respect to that of 300 nm-thick SiO$_2$ [28]. The high efficiency of LG enables the observation of an ambipolar conducting channel in MnPS$_3$. The two-terminal output curves (Fig. 6a) of a MnPS$_3$ EDLT device of 13-nm channel thickness show large current modulations by both positive and negative LG voltages V$_{LG}$. The channel current I$_{ds}$ exhibits super linear dependences on the excitation voltage V$_{ds}$, which is likely caused by a large contact resistance. An enhanced contact quality is likely to improve the device performance. The top-left inset shows the optical image of an EDLT device, and the top-right inset presents the magnified image of the same device. Fig. 6b reveals the transfer curve along with the carrier mobility $\mu = \frac{1}{C}\frac{dG}{dV_{LG}}\frac{L}{W}$ for both electrons and holes, where $C = 8.4\ \mu F cm^{-2}$ is the equivalent capacitance of LG[28] and L and W are the length and width of the conducting channel, respectively. The dramatic increase in channel conductance G when V$_{LG}$ is higher (lower) than the positive (negative) threshold voltage demonstrates the ambipolar conducting operation of our EDLTs with high mobility for electrons (~1.3 cm$^2$ V$^{-1}$ s$^{-1}$) and holes (~2.5 cm$^2$ V$^{-1}$ s$^{-1}$). The inset displays the transfer curve in a logarithmic scale. When V$_{LG}$ varies from +4 V to −4 V, the channel switches from n-type "on" state to "off" state and changed to p-type "on" state with an on/off ratio more than 10$^4$. The on/off ratio is comparable with the values in widely studied TMDCs. The ambipolar performance, as well as the high on/off ratio, render MnPS$_3$ a promising material for low-energy consuming devices with complementary logic. These desirable attributes are crucial to a superb noise margin and robust operation. The creation of an ambipolar conducting channel opens a new avenue for exploring the magnetic order in 2D antiferromagnetic systems. This channel also supplies an ideal terrace for probing valleytronics coupled to antiferromagnetic orders.

In summary, few-layer MnPS$_3$ crystal offers a productive platform for exploring antiferromagnetic order and valley degree of electron freedom in 2D limitation. Fundamental crystal parameters, such as lattice constants, layer-dependent Raman fingerprints and Raman peak anisotropy are measured by HRSTEM and ARPR techniques. Identifying the crystal structure and obtaining the Raman spectrum of MnPS$_3$ flakes lay a solid foundation for follow-up research. The Heisenberg-type antiferromagnetic order in MnPS$_3$ is confirmed by SQUID measurement, critical magnetic properties like effective magnetic momentum and exchange interactions are extracted from MFA model. The confirmation of ambipolar magnetoelectric transport in an antiferromagnetic semiconductor opens a new avenue for exploring fundamental correlated phenomena, i.e. spin/valleytronics coupled to an antiferromagnetic order.

**Experimental section**

**Crystal synthesize**

The MnPS$_3$ single crystals were prepared by a chemical vapor transport method. The stoichiometric mixture of Mn, P, and S powder was sealed in an evacuated quartz tube. Plate-like single crystals can be obtained via the vapor transport growth with a temperature gradient from 650 ℃ to 600 ℃. The composition and structure of
MnPS$_3$ single crystals were checked by X-ray diffraction and Energy-dispersive X-ray spectrometer.

**TEM characterization**

The TEM samples are prepared by direct transfer from scotch tapes to 400-mesh copper grids after mechanical exfoliation. The SAED and EELS are carried out with a JEM 2010F (JEOL, Japan) under 200 kV while the HRSTEM is performed with a JEM ARM 200CF (JEOL, Japan), equipped with a CEOS probe corrector and a cold field-emission gun, also at an acceleration voltage of 200 kV for its highest resolution. A low probe current (less than 75 pA) is chosen to reduce electron radiation and a convergence semiangle of around 35mrad and an inner acquisition semiangle of 79 mrad were used in the HAADF STEM. The HRSTEM image is filtered through the standard Wiener deconvolution to increase its signal-noise ratio for a better display. The zero-loss peak in the EELS data is aligned to exact 0 eV and the Mn-L2,3 edge spectrum go through the power-law background subtraction after that.

**EDLT fabrication**

Thin flakes of MnPS$_3$ are prepared by micro-mechanical exfoliation of a single bulk crystal. The thickness of MnPS$_3$ flakes is verified by atomic force microscopy. Then the flakes are transferred on the top of another prepared boron nitride flake. Electron-beam lithography (EBL) is then applied to define Hall patterns on the MnPS$_3$ flakes followed by electron-beam evaporation and lift-off techniques to deposit contact metals in the defined Hall patterns. Finally standard Hall devices with side gate electrodes are fabricated. The contact metal electrodes consist of Ti/Au/SiO$_2$ (5 nm/60 nm/30 nm) while the side electrodes are not covered by SiO$_2$. The contact metal electrodes are covered by SiO$_2$ to isolate the electrodes from direct contact with the ionic liquid. The sizes of the side-gate electrodes are much larger (area ratio >$10^3$) than those of the $MnPS_3$ flakes. The large area ratio is designed to make sure that the voltage drop is effectively applied at the interface between $MnPS_3$ flakes and the ionic liquid instead of between the side gate electrode and ionic liquid. Thanks to the thin electric layer (~1 nm) formed between the ionic liquid and graphene surface, an extremely strong electric field (~$5\times10^9$ V/m) is generated to introduce an extremely high density of charge carriers to the samples (~$5.23\times10^{13}$ $V^{-1}cm^{-2}$). The ionic liquid is N, N-diethy1-N-(2-methoxyethy1)-N-methyl ammonium bis-(trifluoromethylsulfony1)-imide (DEME-TFSI).

**Acknowledgements**

Financial support from the Research Grants Council of Hong Kong (Project Nos. 16302215, 16300717, GRF16307114 and HKU9/CRF/13G) and technical support of the Raith-HKUST Nanotechnology Laboratory for the electron-beam lithography facility at MCPF are hereby acknowledged.

Work at Tulane University was supported by the U.S. Department of Energy under Grant No. DE-SC0014208 (support for crystal growth).

**References**

1. Lin, M.-W.; Zhuang, H. L.; Yan, J.; Ward, T. Z.; Puretzky, A. A.; Rouleau, C. M.; Gai, Z.; Liang, L.; Meunier, V.; Sumpter, B. G., Ultrathin nanosheets of CrSiTe 3: a semiconducting two-dimensional ferromagnetic material. *Journal of Materials Chemistry C* 2016, *4* (2), 315-322.
2. Tian, Y.; Gray, M. J.; Ji, H.; Cava, R.; Burch, K. S., Magneto-elastic coupling in a potential ferromagnetic 2D atomic crystal. *2D Materials* 2016, *3* (2), 025035.
3. Nicolosi, V.; Chhowalla, M.; Kanatzidis, M. G.; Strano, M. S.; Coleman, J. N., Liquid exfoliation of layered materials. *Science* 2013, *340* (6139), 1226419.
4. Zhang, X.; Zhao, X.; Wu, D.; Jing, Y.; Zhou, Z., MnPSe3 Monolayer: A Promising 2D Visible‐Light Photohydrolytic Catalyst with High Carrier Mobility. *Advanced Science* 2016, *3* (10).
5. Lee, J.-U.; Lee, S.; Ryoo, J. H.; Kang, S.; Kim, T. Y.; Kim, P.; Park, C.-H.; Park, J.-G.; Cheong, H., Ising-Type Magnetic Ordering in Atomically Thin FePS3. *Nano Letters* 2016, *16* (12), 7433-7438.
6. Du, K.-z.; Wang, X.-z.; Liu, Y.; Hu, P.; Utama, M. I. B.; Gan, C. K.; Xiong, Q.; Kloc, C., Weak Van der Waals stacking, wide-range band gap, and Raman study on ultrathin layers of metal phosphorus trichalcogenides. *ACS nano* 2015, *10* (2), 1738-1743.
7. Wang, X.; Du, K.; Liu, Y. Y. F.; Hu, P.; Zhang, J.; Zhang, Q.; Owen, M. H. S.; Lu, X.; Gan, C. K.; Sengupta, P., Raman spectroscopy of atomically thin two-dimensional magnetic iron phosphorus trisulfide (FePS3) crystals. *2D Materials* 2016, *3* (3), 031009.

8. Sandilands, L.; Shen, J.; Chugunov, G.; Zhao, S.; Ono, S.; Ando, Y.; Burch, K., Stability of exfoliated $Bi_2Sr_2Dy_xCa_{1-x}Cu_2O_{8+\delta}$ studied by Raman microscopy. *Physical Review B* 2010, *82* (6), 064503.
9. de Jongh, L. J.; Miedema, A. R., Experiments on simple magnetic model systems. *Advances in Physics* 1974, *23* (1), 1-260.
10. Willett, R. D.; Gatteschi, D.; Kahn, O., Magneto-structural correlations in exchange coupled systems. 1985.
11. González-Herrero, H.; Gómez-Rodríguez, J. M.; Mallet, P.; Moaied, M.; Palacios, J. J.; Salgado, C.; Ugeda, M. M.; Veuillen, J.-Y.; Yndurain, F.; Brihuega, I., Atomic-scale control of graphene magnetism by using hydrogen atoms. *Science* 2016, *352* (6284), 437-441.
12. Nair, R.; Sepioni, M.; Tsai, I.-L.; Lehtinen, O.; Keinonen, J.; Krasheninnikov, A.; Thomson, T.; Geim, A.; Grigorieva, I., Spin-half paramagnetism in graphene induced by point defects. *Nature Physics* 2012, *8* (3), 199-202.
13. McCreary, K. M.; Swartz, A. G.; Han, W.; Fabian, J.; Kawakami, R. K., Magnetic moment formation in graphene detected by scattering of pure spin currents. *Physical review letters* 2012, *109* (18), 186604.
14. Červenka, J.; Katsnelson, M.; Flipse, C., Room-temperature ferromagnetism in graphite driven by two-dimensional networks of point defects. *Nature Physics* 2009, *5* (11), 840-844.
15. Ugeda, M. M.; Brihuega, I.; Guinea, F.; Gómez-Rodríguez, J. M., Missing atom as a source of carbon magnetism. *Physical Review Letters* 2010, *104* (9), 096804.
16. Uchoa, B.; Kotov, V. N.; Peres, N.; Neto, A. C., Localized magnetic states in graphene. *Physical review letters* 2008, *101* (2), 026805.
17. Joy, P.; Vasudevan, S., Magnetism in the layered transition-metal thiophosphates $MPS_3$ (M= Mn, Fe, and Ni). *Physical Review B* 1992, *46* (9), 5425.
18. Mayorga-Martinez, C. C.; Sofer, Z. k.; Sedmidubský, D.; Huber, S. t. p. n.; Eng, A. Y. S.; Pumera, M., Layered Metal Thiophosphite Materials: Magnetic, Electrochemical, and Electronic Properties. *ACS Applied Materials & Interfaces* 2017.
19. Ressouche, E.; Loire, M.; Simonet, V.; Ballou, R.; Stunault, A.; Wildes, A., Magnetoelectric $MnPS_3$ as a candidate for ferrotoroidicity. *Physical Review B* 2010, *82* (10), 100408.
20. Wolf, S.; Awschalom, D.; Buhrman, R.; Daughton, J.; Von Molnar, S.; Roukes, M.; Chtchelkanova, A. Y.; Treger, D., Spintronics: a spin-based electronics vision for the future. *Science* 2001, *294* (5546), 1488-1495.
21. Žutić, I.; Fabian, J.; Sarma, S. D., Spintronics: Fundamentals and applications. *Reviews of modern physics* 2004, *76* (2), 323.
22. Bogani, L.; Wernsdorfer, W., Molecular spintronics using single-molecule magnets. *Nature materials* 2008, *7* (3), 179-186.
23. Awschalom, D. D.; Flatté, M. E., Challenges for semiconductor spintronics. *Nature Physics* 2007, *3* (3), 153-159.
24. Mai, C.; Barrette, A.; Yu, Y.; Semenov, Y. G.; Kim, K. W.; Cao, L.; Gundogdu, K., Many-body effects in valleytronics: Direct measurement of valley lifetimes in single-layer MoS2. *Nano letters* 2013, *14* (1), 202-206.
25. Ezawa, M., Spin valleytronics in silicene: Quantum spin Hall–quantum anomalous Hall insulators and single-valley semimetals. *Physical Review B* 2013, *87* (15), 155415.
26. Nebel, C. E., Valleytronics: Electrons dance in diamond. *Nature materials* 2013, *12* (8), 690-691.
27. Li, X.; Cao, T.; Niu, Q.; Shi, J.; Feng, J., Coupling the valley degree of freedom to antiferromagnetic order. *Proceedings of the National Academy of Sciences* 2013, *110* (10), 3738-3742.

28. Long, G.; Xu, S.; Zhang, T.; Wu, Z.; Wong, W. K.; Han, T.; Lin, J.; Cai, Y.; Wang, N., Charge density wave phase transition on the surface of electrostatically doped multilayer graphene. *Applied Physics Letters* 2016, *109* (18), 183107.
29. Long, G.; Xu, S.; Shen, J.; Hou, J.; Wu, Z.; Han, T.; Lin, J.; Wong, W. K.; Cai, Y.; Lortz, R., Type-controlled nanodevices based on encapsulated few-layer black phosphorus for quantum transport. *2D Materials* 2016, *3* (3), 031001.
30. Long, G.; Maryenko, D.; Shen, J.; Xu, S.; Hou, J.; Wu, Z.; Wong, W. K.; Han, T.; Lin, J.; Cai, Y., Achieving Ultrahigh Carrier Mobility in Two-dimensional Hole Gas of Black Phosphorus. *Nano Letters* 2016.
31. Kurosawa, K.; Saito, S.; Yamaguchi, Y., Neutron diffraction study on mnps3 and feps3. *Journal of the Physical Society of Japan* 1983, *52* (11), 3919-3926.
32. Liu, E.; Fu, Y.; Wang, Y.; Feng, Y.; Liu, H.; Wan, X.; Zhou, W.; Wang, B.; Shao, L.; Ho, C.-H., Integrated digital inverters based on two-dimensional anisotropic ReS2 field-effect transistors. *Nature communications* 2015, *6*.
33. Claassen, H. H.; Selig, H.; Shamir, J., Raman Apparatus Using Laser Excitation and Polarization Measurements. Rotational Spectrum of Fluorine. *Applied Spectroscopy* 1969, *23* (1), 8-12.
34. Allemand, C. D., Depolarization ratio measurements in Raman spectrometry. *Applied Spectroscopy* 1970, *24* (3), 348-353.
35. Gonze, X.; Jollet, F.; Araujo, F. A.; Adams, D.; Amadon, B.; Applencourt, T.; Audouze, C.; Beuken, J.-M.; Bieder, J.; Bokhanchuk, A., Recent developments in the ABINIT software package. *Computer Physics Communications* 2016, *205*, 106-131.
36. Gonze, X.; Amadon, B.; Anglade, P.-M.; Beuken, J.-M.; Bottin, F.; Boulanger, P.; Bruneval, F.; Caliste, D.; Caracas, R.; Côté, M., ABINIT: First-principles approach to material and nanosystem properties. *Computer Physics Communications* 2009, *180* (12), 2582-2615.
37. Gonze, X., A brief introduction to the ABINIT software package. *Zeitschrift für Kristallographie-Crystalline Materials* 2005, *220* (5/6), 558-562.
38. Gonze, X.; Beuken, J.-M.; Caracas, R.; Detraux, F.; Fuchs, M.; Rignanese, G.-M.; Sindic, L.; Verstraete, M.; Zerah, G.; Jollet, F., First-principles computation of material properties: the ABINIT software project. *Computational Materials Science* 2002, *25* (3), 478-492.
39. Wu, J.; Mao, N.; Xie, L.; Xu, H.; Zhang, J., Identifying the Crystalline Orientation of Black Phosphorus Using Angle‐Resolved Polarized Raman Spectroscopy. *Angewandte Chemie* 2015, *127* (8), 2396-2399.
40. Okuda, K.; Kurosawa, K.; Saito, S.; Honda, M.; Yu, Z.; Date, M., Magnetic properties of layered compound mnps3. *Journal of the Physical Society of Japan* 1986, *55* (12), 4456-4463.
41. Sales, B.; Wohlleben, D., Susceptibility of interconfiguration-fluctuation compounds. *Physical Review Letters* 1975, *35* (18), 1240.
42. Hamaker, H.; Woolf, L.; MacKay, H.; Fisk, Z.; Maple, M., Coexistence of superconductivity and antiferromagnetic order in SmRh4B4. *Solid State Communications* 1979, *32* (4), 289-294.
43. Tan, H.; Verbeeck, J.; Abakumov, A.; Van Tendeloo, G., Oxidation state and chemical shift investigation in transition metal oxides by EELS. *Ultramicroscopy* 2012, *116*, 24-33.
44. Breed, D., Antiferromagnetism of K2MnF4. *Physics Letters* 1966, *23* (3), 181-182.

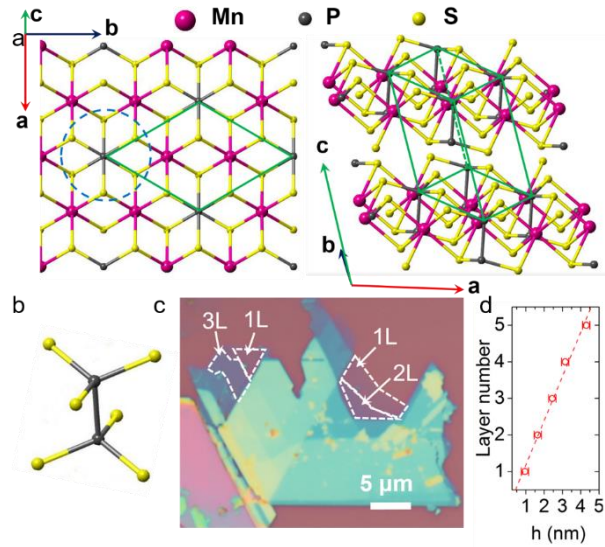

**Fig. 1 Crystal structure of MnPS$_3$.** (a) Ball-and-stick model of the MnPS$_3$ crystal structure. The left panel shows the top view of the monolayer model. The green parallelogram indicates the primary unit cell. The arrows show the crystalline orientations (red: a; blue: b; greed: c). The right panel displays the side view of the bilayer model, and the green parallelepiped indicates the primary unit cell of the MnPS$_3$ bulk crystal. (b) The dumbbell-shaped structure of the [P$_2$S$_6$]$^{4-}$ unit is marked by the dashed light blue circle in (a). (c) Micro-optical image of an exfoliated few-layer MnPS$_3$. The scale bar denotes 5 μm. (d) Measured thicknesses $h$ and corresponding layer numbers. The red dashed line represents the linear fitting result.

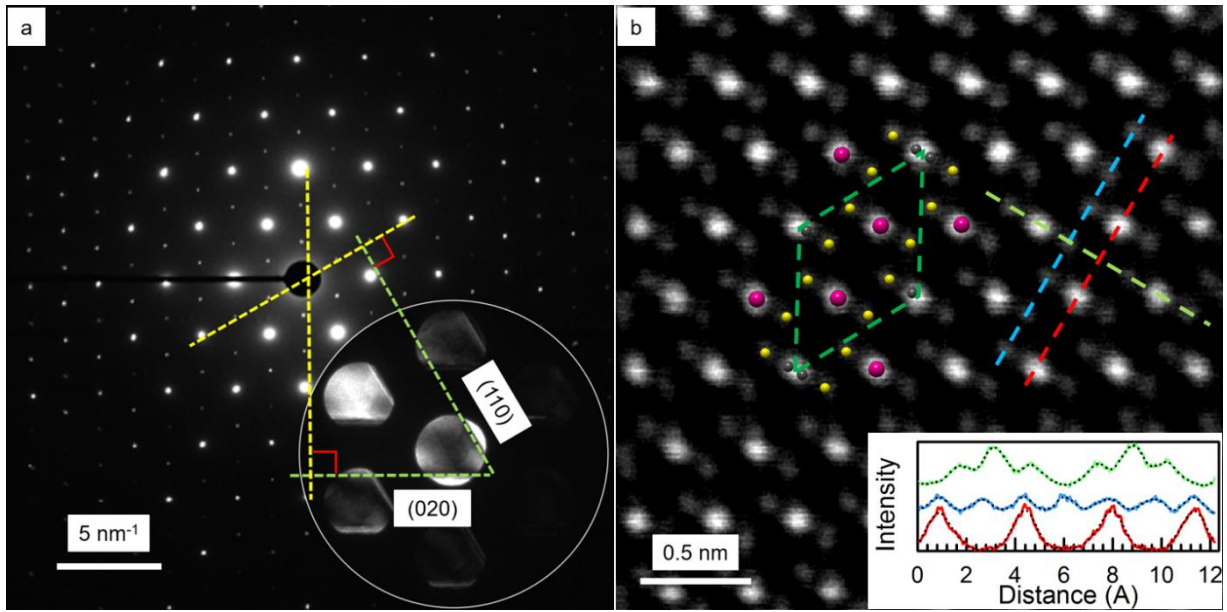

**Fig. 2 Transmission electron microscopy (TEM) images of MnPS₃.** (a) SAED pattern of MnPS₃. The scale bar represents 5 nm$^{-1}$. The inset shows the defocused diffraction pattern at the same site. The real space features are apparent in the defocused transmitted spot. The light green dashed lines indicate the two edges of the MnPS₃ sample, whereas the yellow dashed lines show the directions perpendicular to the two edges. A magnetic declination of 3° is corrected. The two edges are indexed to be (110) and (020) according to the diffraction patterns. (b) HAADF STEM image of MnPS₃ sample along c direction. The scale bar represents 0.5 nm. The colored balls indicate the different elements in the image (purple: Mn; black: P; yellow: S). The dashed green parallelogram denotes the primary unit cell. The inset indicates that the intensities vary with distance along the directions, as marked by the dashed lines with the same colors. A constant offset between different lines is introduced for a clear display.

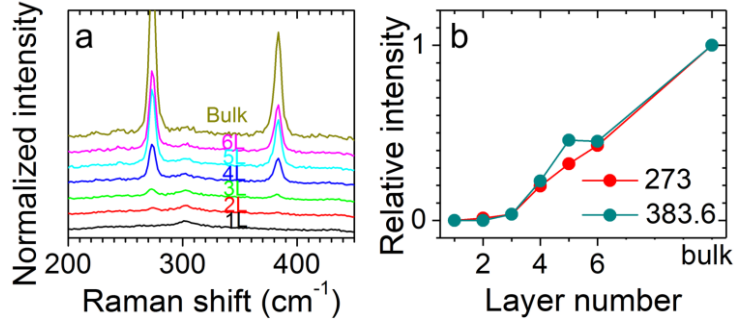

**Fig. 3 Thickness-dependent Raman spectrum of MnPS$_3$.** (a) Raman spectra of MnPS$_3$ with different thicknesses (monolayer to bulk). The Raman intensities are normalized to the peak intensity of substrate Si. (b) MnPS$_3$ Raman peak intensities change with thickness. The peak intensities of the bulk sample is regarded as one unit.

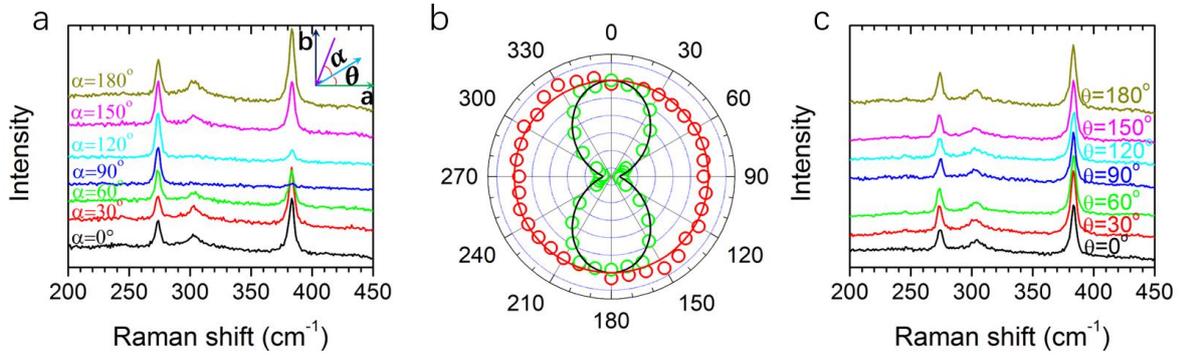

**Fig. 4 Angle-resolved polarized Raman spectrum of MnPS$_3$.** (a) Polarized Raman spectra of MnPS$_3$ with $\theta = 0$. The top-right inset shows the ARPR spectrum measurement configuration. The green and blue arrows indicate the a and b directions, respectively. The purple and light blue arrows indicate the linear polarization directions of the incident and scattered lights, respectively. Angle $\alpha$ is the angle between the incident and scattered lights, whereas angle $\theta$ is the angle between the scattered light and the *a* direction. (b) $\alpha$-dependent polarized Raman peak intensities with $\theta = 0$ (Red: P$_{283}$; green: P$_{383}$). The peak intensities at $\alpha = 0$ is considered as one unit. The red and black solid lines display the fitting results of constant intensity and $\cos^2(\alpha)$, respectively. (c) $\theta$-dependent polarized Raman spectrum with $\alpha = 0$.

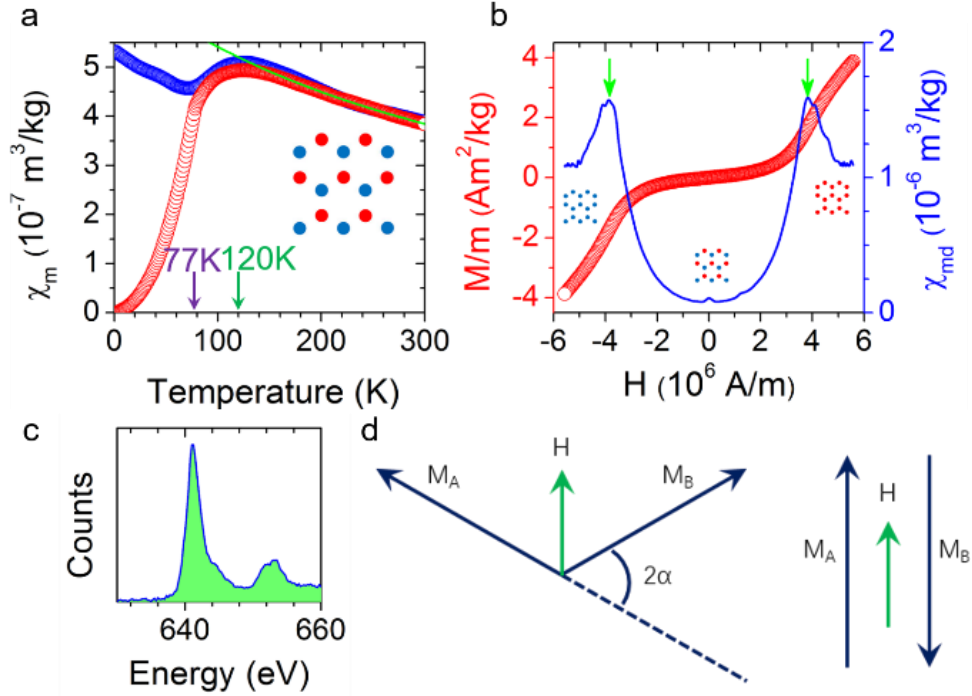

**Fig. 5 Magnetic susceptibility of MnPS$_3$.** (a) Temperature dependences of mass magnetic susceptibility $\chi_m$ with in-plane (blue) and out-plane (red) magnetic fields of 1600 A/m. The green line shows the fitting result of Curie-Weiss formula. The purple arrow indicate the critical temperature between paramagnetic and antiferromagnetic orders. The green arrow indicate the temperature corresponding to the peak value of $\chi_m$. The inset displays the magnetic order of MnPS$_3$ in antiferromagnetic order regime (blue: spin up; red: spin down; or inverse). (b) Mass magnetization (red) and derivate mass susceptibility (blue) as functions of out-plane magnetic field at T=5K. The three inset patterns display the magnetic orders at magnetic field ranges (blue: spin up; red: spin down; or inverse). The green arrows indicate the peak positions of derivate mass susceptibility. (c) Electron energy loss spectrum (EELS) of the few-layer sample shows characteristic peaks from [Mn]$^{2+}$. (d) Mean field approximation (MFA) model for in- plane (left) and out-plane (right) field magnetic susceptibility in antiferromagnetic regime. M$_A$ and M$_B$ indicate the magnetizations with opposite orientations, H is the applied magnetic field.

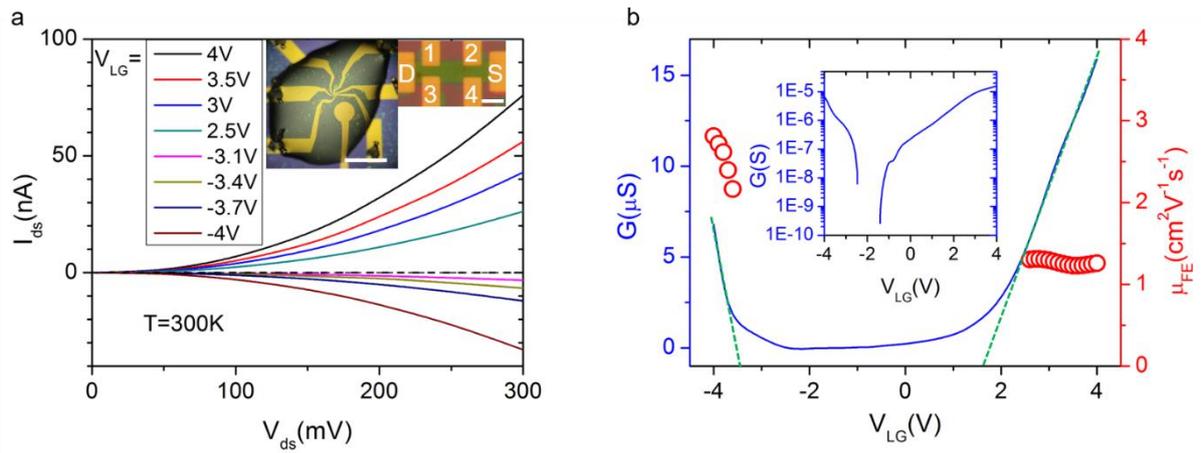

**Fig. 6 Transport features of electrical double layer transistor (EDLT) based on MnPS$_3$.** (a) Output curves of MnPS$_3$ at different LG voltages. Current of electrons (positive LG voltage) are shown with positive indices, and those of electrons (negative LG voltage) are shown with negative indices. The two insets show the micro-optical images of the EDLT in different ratios. The scale bars denote 300 μm (left) and 3 μm (right). (b) Room temperature transfer characteristics (blue line) and the extracted field-effect mobility as a function of $V_{LG}$ (red circles) of the MnPS$_3$ EDLT. The green dashed lines show the linear fitting result of transfer characteristics. The inset shows the transfer characteristics in the log scale.

TOC:

**Raman spectroscopy** as long as the electronic transport properties of the hexagonal antiferromagnetic lattice in MnPS$_3$ is studied. Raman fingerprints of MnPS$_3$ with different thicknesses are obtained. Raman spectrum evolution of when cooling down to cryogenic temperature reveals the universal magnetic orders in MnPS$_3$. A liquid gating technique is applied to probe the transport properties of MnPS$_3$.

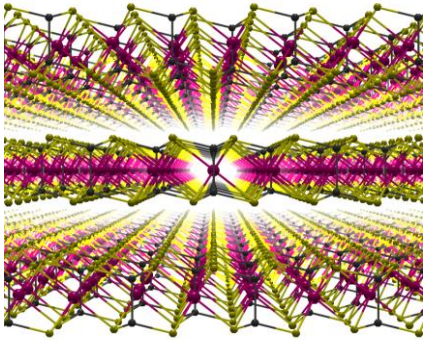